\documentclass[showkeys,aps,amssymb,floatfix,prd]{revtex4}
\usepackage{bm,graphicx}

\newcommand\beq{\begin{equation}}
\newcommand\beqa{\begin{eqnarray}}
\newcommand\beqan{\begin{eqnarray*}}
\newcommand\eeq{\end{equation}}
\newcommand\eeqa{\end{eqnarray}}
\newcommand\eeqan{\end{eqnarray*}}

\newcommand\Mbh{M_\bullet}
\newcommand\gravr{{\sf m}_\bullet}

\newcommand\vthE{\vartheta_E}
\newcommand\vep{\varepsilon}
\newcommand\vth{\vartheta}
\newcommand\vthbh{\vartheta_\bullet}
\newcommand\cb{{\cal B}}
\newcommand\tauhat{\hat{\tau}}
\newcommand\order[2]{\mathcal{O}\left({#1}\right)^{#2}}
\newcommand\assump[1]{[A#1]}

\newcommand\E[1]{\times10^{#1}}

\newcommand\reffig[1]{Fig.~\ref{fig:#1}}

\begin{document}

\title{Formalism for testing theories of gravity using lensing by \\
compact objects. II: Probing Post-Post-Newtonian metrics}

\author{Charles R. Keeton}
\affiliation{Department of Physics \& Astronomy, Rutgers University,
  136 Frelinghuysen Road, Piscataway, NJ 08854;\\
  {\tt keeton@physics.rutgers.edu}}

\author{A. O. Petters}
\affiliation{Departments of Mathematics and Physics,
  Duke University,\\
  Science Drive,\\ Durham, NC 27708-0320;\\
  {\tt petters@math.duke.edu}\\
\\
{\rm Submitted January 13, 2006}
}

\begin{abstract}
We study gravitational lensing by compact objects in gravity
theories that can be written in a Post-Post-Newtonian (PPN)
framework: i.e., the metric is static and spherically symmetric,
and can be written as a Taylor series in $\gravr/r$, where
$\gravr$ is the gravitational radius of the compact object.
Working invariantly, we compute corrections to standard
weak-deflection lensing observables at first and second order
in the perturbation parameter $\vep = \vthbh/\vthE$, where
$\vthbh$ is the angular gravitational radius and $\vthE$ is
the angular Einstein ring radius of the lens.  We show that
the first-order corrections to the total magnification and
centroid position vanish universally for gravity theories
that can be written in the PPN framework.  This arises from
some surprising, fundamental relations among the lensing
observables in PPN gravity models.  We derive these relations
for the image positions, magnifications, and time delays.
A deep consequence is that any violation of the universal
relations would signal the need for a gravity model outside
the PPN framework (provided that some basic assumptions hold).
In practical terms, the relations will guide observational
programs to test general relativity, modified gravity theories,
and possibly the Cosmic Censorship conjecture.  We use the
new relations to identify lensing observables that are
accessible to current or near-future technology, and to find
combinations of observables that are most useful for probing
the spacetime metric.  We give explicit applications to the
Galactic black hole, microlensing, and the binary pulsar
J0737$-$3039.
\end{abstract}


\maketitle

\section{Introduction}
\label{sec:intro}

Gravitational lensing is now a central field of astronomy with
wide-ranging applications that relate to extra-solar planets,
dark matter substructures, and cosmological parameters
(including dark energy) \cite{SEF,petters,Saas-Fee}.
Theoretical studies have also explored lensing by black
holes within the context of
general relativity \cite{darwin}-\cite{PPN-2},
braneworld gravity \cite{bhadra}-\cite{majumdar},
and string theory \cite{gibbons,paperI}.
Most black hole lensing studies have focused on
``relativistic'' images corresponding to light rays that loop
around a black hole, which probe gravity in the strong-deflection
limit, but are extremely difficult to detect observationally
\cite{virbhadra,petters-SgrA}.  We are proposing and evaluating
new possibilities for using gravitational lensing by compact
objects (including black holes) to test various theories of
gravity using current or near-future technology.

In Paper I of this series \cite{paperI}, we introduced an
analytic framework for studying gravitational lensing by a
compact deflector with mass $\Mbh$, in which the lensing
scenario satisfies three basic assumptions:
\begin{itemize}

\item[\assump{1}] The gravitational lens is compact, static,
and spherically symmetric, with an asymptotically flat spacetime
geometry sufficiently far from the lens \footnote{In lensing,
asymptotic flatness can be generalized to a Robertson-Walker
background by using angular diameter distances and including
appropriate redshift factors.}.  The spacetime is vacuum
outside the lens and flat in the absence of the lens.

\item[\assump{2}] The observer and source lie in the
asymptotically flat regime of the spacetime.

\item[\assump{3}] The light ray's distance of closest approach
$r_0$ and impact parameter $b$ both lie well outside the
gravitational radius $\gravr = G \Mbh/c^2$, namely,
$\gravr/r_0 \ll 1$ and $\gravr/b \ll 1$.  The bending angle
can then be expressed as a series expansion in $\gravr/b$, as
follows:
\beq \label{eq:bangle-ser}
  \hat{\alpha}(b) = A_1 \left(\frac{\gravr}{b}\right)
    \ + \ A_2 \left(\frac{\gravr}{b}\right)^2
    \ + \ A_3 \left(\frac{\gravr}{b}\right)^3
    \ + \ \order{\frac{\gravr}{b}}{4} .
\eeq
The coefficients $A_i$ are independent of $\gravr/b$, but may
include other fixed parameters of the spacetime.  Since $b$ and
$\gravr$ are invariants of the light ray, (\ref{eq:bangle-ser})
is independent of coordinates.  Note that the subscript of $A_i$
conveniently indicates that the component is affiliated with a
term of order $i$ in $\gravr/b$.

\end{itemize}
We applied the lensing framework to gravity theories that can
be written in a Post-Post-Newtonian (PPN) expansion, meaning
that the metric can be written as a Taylor series in $\gravr/r$.
Working invariantly, we computed the observable properties of
the primary (positive-parity) and secondary (negative-parity)
lensed images.  Assuming that \assump{3} holds, we wrote the
lensing observables as series expansions in $\vep = \vthbh/\vthE$,
where $\vthbh$ is the angular gravitational radius and $\vthE$
is the angular Einstein ring radius.  For gravity theories that
agree with general relativity in the weak-deflection limit
($A_1 = 4$ in eqn.~\ref{eq:bangle-ser}), the zeroth-order terms
in the $\vep$ expansions give the familiar results for lensing
by a point mass in the weak-deflection limit of general
relativity.  The higher order terms give corrections to the
lensing observables, which differ for different gravity theories.
We studied general third-order PPN models, which allowed us to
compute the weak-deflection lensing results plus the first two
correction terms (order $\vep$ and $\vep^2$).

In Paper I we found the interesting result that the
first-order corrections to two lensing observables --- the
total magnification and the magnification-weighted centroid
position --- vanish in PPN gravity models that agree with
general relativity in the weak-deflection limit (i.e.,
$A_1 = 4$).  More generally, we found that these corrections
depend on $A_1-4$, suggesting that they nearly vanish in
gravity theories that agree only approximately with general
relativity in the weak-deflection regime (i.e., $A_1 \approx 4$).
Note that existing observations constrain this parameter to
$A_1 = 3.99966 \pm 0.00090$ \cite{sh}.

Working more generally than in Paper I, we have now
discovered the surprising result that the first-order
corrections to the total magnification and centroid in fact
vanish exactly in all PPN models.  This depends on precise
cancellations that are striking because the PPN framework
covers quite a broad range of gravity theories.  Understanding
why the cancellations occur has led us to identify some new
fundamental relations between lensing observables in the
PPN framework.  Some of the relations are universal for
PPN models, in the sense that they hold for all values of
the invariant PPN parameters of the light bending angle.

These new lensing relations will play key roles in planning
observing missions to test theories of gravity.  First, the
relations allow us to determine which observables are most
accessible to current or near-future instrumentation.  We
give explicit applications to the Galactic black hole,
Galactic microlensing, and the binary pulsar J0737-3039.
Second, the lensing relations help us identify combinations
of lensing observables that are most useful for probing the
spacetime metric by constraining invariant PPN parameters.  
One of the measurable parameters is connected to the
existence of naked singularities in certain gravity models
(see Paper I), so we may have an observational test of the
Cosmic Censorship conjecture \cite{CosmicCensorship}.
Third, the universal relations provide a powerful means to
test the entire PPN framework.  Any violation of relations
that are universal in the PPN framework would suggest that
a fundamentally different theory of gravity is needed
(provided that assumptions \assump{1}--\assump{3} hold).

In this paper we present the new lensing relations and use them
to assess prospects for using lensing observations to test PPN
gravity theories.  Section~\ref{sec:formalism} reviews the
third-order PPN lensing framework.  Section~\ref{sec:relations}
derives the new relations between the image positions,
magnifications, and time delays, and discusses their conceptual
implications.  Section~\ref{sec:applications} considers
applications of the relations to various astrophysical settings.

\section{Lensing in the PPN Framework}
\label{sec:formalism}

In this section we review our results for lensing in the PPN
framework.  See Paper I for the complete analysis.

Consider a compact body of mass $\Mbh$, perhaps a black hole
or neutron star, that is described by a geometric theory of
gravity.  By assumptions \assump{1}--\assump{3}, it suffices
to analyze an equatorial metric of the form
\beq \label{eq:metric-equatorial}
  ds^2 = - A(r)\,dt^2 \ + \ B(r)\,dr^2 \ + \ r^2\,d\varphi^2\,.
\eeq
We study metrics whose coefficients can be expressed in
third-order PPN expansions,
\beqa
A(r) &=& 1 \ + \ 2\, a_1 \, \left(\frac{\phi}{c^2}\right)
  \ + \ 2\, a_2\, \left(\frac{\phi}{c^2}\right)^2
  \ + \ 2\, a_3\, \left(\frac{\phi}{c^2}\right)^3 
  \ + \ \ldots\,, \label{eq:PPN-A} \\
B(r) &=& 1 \ - \ 2\, b_1 \, \left(\frac{\phi}{c^2}\right)
  \ + \ 4\, b_2\, \left(\frac{\phi}{c^2}\right)^2
  \ - \ 8\, b_3\, \left(\frac{\phi}{c^2}\right)^3
  \ + \ \ldots\,, \label{eq:PPN-B}
\eeqa
where $\phi$ is the three-dimensional Newtonian potential with
\beq
\frac{\phi}{c^2} = - \frac{\gravr}{r}
\eeq

Section III.C of Paper I derives the light bending angle for
this metric.  The invariant expression for the bending angle
takes the form of (\ref{eq:bangle-ser}) with coefficients
\beqa
  A_1 &=& 2(a_1+b_1)\,,
\label{eq:PPN-A1}\\
  A_2 &=& \left( 2 a_1^2 - a_2 + a_1 b_1 - \frac{b_1^2}{4} + b_2 \right) \pi\,,
\label{eq:PPN-A2}\\
  A_3 &=& \frac{2}{3} \Bigl[ 35 a_1^3 + 15 a_1^2 b_1
    - 3 a_1 \left(10 a_2 + b_1^2 - 4 b_2 \right)
    + 6 a_3 + b_1^3 - 6 a_2 b_1 - 4 b_1 b_2 + 8 b_3 \Bigr] .
\label{eq:PPN-A3}
\eeqa
The coordinate independent quantities $A_i$ will be called the
{\em invariant PPN parameters} of the light bending angle.
For reference, the Schwarzschild metric in general relativity
has $a_1 = b_1 = b_2 = b_3 = 1$ and $a_2 = a_3 = 0$, and hence
$A_1 = 4$, $A_2 = 15\pi/4$, and $A_3 = 128/3$.

\begin{figure}
\includegraphics[width=5.0in]{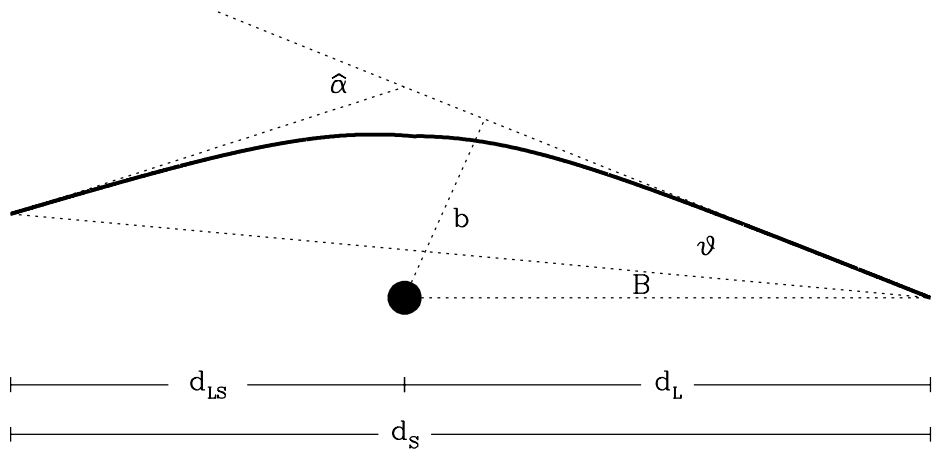}
\caption{
Schematic diagram of the lensing geometry.  Standard quantities
are identified:
$\cb$ is the angular position of the unlensed source;
$\vth$ is the angular position of an image;
$\hat{\alpha}$ is the bending angle;
and $d_L$, $d_S$, and $d_{LS}$ are angular diameter distances
between the observer, lens, and source.  The impact parameter
$b$ is an invariant of the light ray and is related to the
angular image position by $\vth = \sin^{-1}(b/d_L)$.
}
\label{fig:geom}
\end{figure}

\reffig{geom} displays the gravitational lensing scenario.
Elementary trigonometry establishes the relationship (see
\cite{virbhadra})
\beq \label{eq:lebh-1}
  \tan \cb = \tan \vth - D \ (\tan \vth + \tan (\hat{\alpha} - \vth))\,,
\eeq
where $D = d_{LS}/d_S$.  This equation agrees well with
the full relativistic formalism for light propagation
\cite{frittelli}, so we take it as the general form of the
lens equation.  The equation has two primary (weak-deflection)
solutions \footnote{If the lens is a black hole, there is also
a countably infinite set of ``relativistic'' images very close
to the black hole's photon sphere, corresponding to light rays
that loop around the lens $1,2,3,\ldots$ times
\cite{darwin}-\cite{virbhadra}, but they are very faint and
we do not consider them further.}: one corresponding to an
image on the same side of the lens as the source; and one on
the opposite side.  By convention, angles describing image
positions are taken to be positive.  This forces the source's
angular position to have different signs: $\cb$ is positive
when we are studying an image on the same side of the lens as
the source (as depicted in \reffig{geom}); while $\cb$ is
negative when we are studying an image on the opposite of the
lens from the source.

Following Paper I, we convert to scaled variables
\beq \label{eq:newvar}
  \beta = \frac{\cb}{\vthE}\ , \quad
  \theta = \frac{\vth}{\vthE}\ , \quad
  \tauhat = \frac{\tau}{\tau_E}\ , \quad
  \vep = \frac{\vthbh}{\vthE} = \frac{\vthE}{4\,D}\, ,
\eeq
where $\vthbh = \tan^{-1}(\gravr/d_L)$ is the angle
subtended by the gravitational radius, and $\tau$ is the
lensing time delay.  The natural angular scale is given by
the angular Einstein ring radius,
\beq \label{eq:thetaE}
  \vthE = \sqrt{\frac{4 G \Mbh d_{LS}}{c^2 d_L d_S}}\ ,
\eeq
while the natural time scale is
\beq
  \tau_E \equiv \frac{d_L\,d_S}{c\,d_{LS}}\ \vthE^2
  = 4\ \frac{\gravr}{c}\ .
\eeq

We then assume that solutions of the lens equation can be
written as a series of the form
\beq \label{eq:tseries}
  \theta = \theta_0 + \theta_1\,\vep + \theta_2\,\vep^2 + 
    \order{\vep}{3} ,
\eeq
where $\theta_0$ represents the image position in the
weak-deflection limit, while $\theta_1$ and $\theta_2$ give
the first- and second-order correction terms.  With these
substitutions, the lens equation becomes
\beqa
  0 &=& D \left[ -4\beta + 4\theta_0 - \frac{A_1}{\theta_0} \right] \vep
\ + \ \frac{D}{\theta_0^2} \Biggl[ -A_2
    + \left( A_1+4\theta_0^2 \right) \theta_1 \Biggr]\,\vep^2 \nonumber\\
&&
\ + \ \frac{D}{3\theta_0^3} \Biggl[ - A_1^3 - 3 A_3 + 12 A_1^2 D \theta_0^2
    - A_1 (56 D^2 \theta_0^4 + 3 \theta_1^2 - 3 \theta_0 \theta_2) \nonumber\\
&&
\qquad\qquad
    + 64 D^2 \theta_0^3 (\theta_0^3-\beta^3) + 6 A_2 \theta_1
    + 12 \theta_0^3 \theta_2 \Biggr]\,\vep^3
\ + \ \order{\vep}{4} . \label{eq:lebh-2}
\eeqa
We solve for $\theta_0$, $\theta_1$, and $\theta_2$ by finding
the values that make each term of the lens equation vanish.
From the vanishing of the first term we obtain the relation
\beq \label{eq:lens0}
  \beta = \theta_0 - \frac{A_1}{4 \theta_0}\ .
\eeq
This is the generalization of the familiar weak-deflection
lens equation for a point mass to $A_1 \ne 4$.  Its solution
is
\beq \label{eq:theta0}
  \theta_0 = \frac{1}{2}\left(\sqrt{ A_1 + \beta^2} + \beta \right) .
\eeq
Then requiring that the second and third terms in
(\ref{eq:lebh-2}) vanish yields the correction terms
\beqa
  \theta_1 &=& \frac{A_2}{A_1+4\theta_0^2}\ ,
    \label{eq:theta1} \\
  \theta_2 &=& \frac{1}{3\,\theta_0\,(A_1+4\theta_0^2)^3} \Biggl[
    A_1 \Bigl(  - 3 A_2^2 + 3 A_1 A_3 - A_1^4 (D^2 - 1)\Bigr)
    \label{eq:theta2} \\
  && \ + \ 4 \Bigl( - 6 A_2^2 +  6 A_1 A_3 +  A_1^4 (D -2 )(D -1) 
    \Bigr) \theta_0^2 \nonumber\\
  && \ + \ 8 \Bigl( 6 A_3  + A_1^3 ( 2 + D (11 D - 12)) \Bigr) \theta_0^4 
    \nonumber\\
  && \ + \ 64 A_1^2 D (4 D - 3)\ \theta_0^6 
     \ + \ 128 A_1 D^2 \ \theta_0^8 \Biggr] . \nonumber
\eeqa
Notice that the $\order{\vep}{2}$ term in the lens equation
(\ref{eq:lebh-2}) does not explicitly involve $\beta$; so
$\theta_1$ depends on the source position only implicitly
through $\theta_0$.  By contrast, the $\order{\vep}{3}$ term
in the lens equation does involve $\beta$.  In writing the
expression for $\theta_2$, we have found it convenient to
substitute for $\beta$ using (\ref{eq:lens0}).  Then the two
correction terms $\theta_1$ and $\theta_2$ are both written
only in terms of $\theta_0$.  The source position dependence
could be made explicit by substituting for $\theta_0$ using
(\ref{eq:theta0}).

The signed magnification $\mu$ of a lensed image at angular
position $\vth$ is given by
\beq
  \mu(\vth) = \left[\frac{\sin \cb (\vth)}{\sin \vth} \ 
    \frac{d \cb(\vth)}{d \vth} \right]^{-1} .
\eeq
After taking the derivative, we change to our scaled angular
variables from (\ref{eq:newvar}) and (\ref{eq:tseries}), and
substitute for $\theta_1$ and $\theta_2$ using
(\ref{eq:theta1})--(\ref{eq:theta2}).  This yields a series
expansion for the magnification,
\beq \label{eq:muser}
  \mu = \mu_0 \ + \ \mu_1\,\vep \ + \ \mu_2\,\vep^2
    \ + \ \order{\vep}{3} ,
\eeq
where (see eqs.~77--79 of Paper I)
\beqa
  \mu_0 &=& \frac{16\theta_0^4}{16\theta_0^4-A_1^2}\ ,
    \label{eq:mu0} \\
  \mu_1 &=& - \frac{16 A_2 \theta_0^3}{(A_1+4\theta_0^2)^3}\ ,
    \label{eq:mu1} \\
  \mu_2 &=& \frac{8\theta_0^2}{3 (A_1-4\theta_0^2) (A_1+4\theta_0^2)^5}
    \Biggl[ - A_1^6 D^2 + 8 A_1^2  \Bigl( 6  A_3  +  A_1^3 ( 2 + 6D - 9D^2)
    \Bigr) \ \theta_0^2
    \label{eq:mu2} \\
  &&\quad - \ 32 \Bigl( 18 A_2^2 - 12 A_1 A_3
    +  A_1^4 \left( D (17 D - 12) - 4 \right)
    \Bigr)\ \theta_0^4  \nonumber\\
  &&\quad + \ 128\, \Bigl( 6 A_3 + A_1^3 ( 2 + 6 D - 9 D^2) \Bigr)\ \theta_0^6 
    - \ 256\, A^2 D^2 \theta_0^8 \ \Biggr] . \nonumber
\eeqa
We have again substituted for $\beta$ using (\ref{eq:lens0}).

The time delay (relative to an undeflected light ray) was
derived in Section V.C of Paper I.  It can be written as a
series of the form
\beq \label{eq:tau}
  \tauhat = \tauhat_0 \ + \ \tauhat_1\,\vep \ + \ \order{\vep}{2} ,
\eeq
where
\beqa
  \tauhat_0 &=& \frac{1}{2} \left[ a_1 + \beta^2 - \theta_0^2
    - \frac{A_1}{4}\ \ln\left(\frac{d_L \theta_0^2 \vthE^2}{4 d_{LS}}\right)
    \right] , \label{eq:tau0} \\
  \tauhat_1 &=& \frac{A_2}{4 \theta_0}\ . \label{eq:tau1}
\eeqa
It is possible to derive the second-order correction to the
time delay, but we have found that to be less important than
the second-order corrections to the position and magnification.

{\em Remark.}  In Paper I, we gave the relation between $\beta$
and $\theta_0$ only for the case $A_1 = 4$.  As a result, there
are minor changes in the expressions for $\theta_2$ and $\mu_2$
in Paper I when $A_1 \ne 4$.  Equations (\ref{eq:theta0}),
(\ref{eq:theta2}), and (\ref{eq:mu2}) give the correct
expressions for the general case in which $A_1$ can take on
any value.  Note, however, that observationally
$A_1 = 3.99966 \pm 0.00090$ \cite{sh}.

\section{New Relations Between Lensing Observables}
\label{sec:relations}

In this section we uncover several new relations between the
perturbation coefficients of the fundamental lensing observables,
namely the positions, magnifications, and time delays of the
two primary images.  Some of the relations are universal among
PPN models in the sense that they hold for any values of the
PPN parameters.  Others hold for all source positions.

\subsubsection{Position relations}

Starting from (\ref{eq:theta0}) and recalling that the
positive- and negative-parity images correspond to $\beta>0$
and $\beta<0$, respectively, we can write the positions of
the two images in the weak-deflection limit as
\beq \label{eq:theta0pm}
  \theta_0^\pm = \frac{1}{2}\left(\sqrt{ A_1 + \beta^2} \pm |\beta| \right) .
\eeq
We can immediately identify two interesting relations:
\beqa
  \theta_0^{+} \ - \ \theta_0^{-} &=& |\beta|\,, \label{eq:imgreln-0} \\
  \theta_0^{+}\ \theta_0^{-} &=& \frac{A_1}{4}\ . \label{eq:tptm=1}
\eeqa
Equation (\ref{eq:imgreln-0}) represents our first universal
relation.  This relation is familiar from standard weak-deflection
lensing in general relativity (see p.~189 of \cite{petters}), but
now we see that it holds for all PPN models, regardless of whether
they agree with general relativity in the weak-deflection limit.
The second equation is our first example of a relation that is
independent of the source position.

Next, combining (\ref{eq:theta1})--(\ref{eq:theta2}) and
(\ref{eq:tptm=1}) yields
\beqa
  \theta_1^+ \ + \ \theta_1^- &=& \frac{A_2}{A_1}\ , \label{eq:imgreln-1} \\
  \theta_1^+ \ - \ \theta_1^- &=& - \frac{A_2\, |\beta|}
    {A_1\, \sqrt{A_1 + \beta^2}}\ , \\
  \theta_2^+ \ - \ \theta_2^- &=& \frac{2 |\beta|}{3 A_1^3}
    \Bigl[ 6 A_2^2 - 6 A_1 A_3 - A_1^4 (2 - 3 D^2) \Bigr] .
    \label{eq:imgreln-2}
\eeqa
The first-order relation for $\theta_1^+ + \theta_1^-$
is another source-independent relation, while the one for
$\theta_1^+ - \theta_1^-$ is source dependent.  Both
first-order relations depend on the sign of $A_2$.  The
second-order relation is independent of the sign of $A_2$.
The dependence on the sign of $A_2$ is interesting because
in Paper I we found that in certain gravity theories this
sign is connected to the occurrence of naked singularities.

\subsubsection{Magnification relations}

We take the magnification terms (\ref{eq:mu0})--(\ref{eq:mu2})
and write $\theta_0$ in terms of $\beta$ using (\ref{eq:theta0pm}).
This yields
\beqa
  \mu_0^{\pm} &=& \frac{1}{2} \pm \frac{A_1 + 2 \beta^2}{4 |\beta|
    \sqrt{A_1+\beta^2}}\ , \label{eq:mu0-beta} \\
  \mu_1^{\pm} &=& - \frac{A_2}{4 (A_1+\beta^2)^{3/2}}\ , \label{eq:mu1-beta} \\
  \mu_2^{\pm} &=& \pm \frac{ 9 A_2^2 + 2(A_1+\beta^2)
    \{-6 A_3 + 2 A_1^2 D^2 \beta^2 + A_1^3 [-2+3D(-2+3D)]\} }
    { 24 |\beta| (A_1+\beta^2)^{5/2} }\ . \label{eq:mu2-beta}
\eeqa
We can then identify three universal magnification relations:
\beq \label{eq:magreln}
  \mu_0^+  \ +  \ \mu_0^- = 1, \qquad
  \mu_1^+  \ - \  \mu_1^- = 0, \qquad 
  \mu_2^+  \ +  \ \mu_2^- = 0.
\eeq
Recall that the sign of the magnification indicates the parity
of an image, so $|\mu|$ actually gives the image brightness.
The zeroth-order relation can be rewritten as
$|\mu_0^+| - |\mu_0^-| = 1$.  In other words, in the PPN
framework the difference between the fluxes of the images (at
zeroth order) always equals the flux of the source in the
absence of lensing. 	

The first-order magnification relation arises because the
first-order correction term $\mu_1$ is the same for both images,
but the actual magnifications have opposite signs.  If $A_2$
is positive, then $\mu_1^{\pm} < 0$.  This makes the
positive-parity image less positive, or fainter; but it makes
the negative-parity image more negative, or brighter.  (If
$A_2$ is negative, the opposite occurs.)  Consequently, the
magnifications of the positive- and negative-parity images
are shifted by the same amount but in the opposite sense.

In addition, combining (\ref{eq:theta0})--(\ref{eq:theta1})
with (\ref{eq:mu1})--(\ref{eq:mu2}) yields another universal
relation,
\beq \label{eq:posmag}
  (\mu_0^+ \, \theta_1^+  \ + \ \mu_0^- \, \theta_1^-)
  \ + \ (\mu_1^+\, \theta_0^+ \ + \ \mu_1^-\, \theta_0^-) = 0.
\eeq
This relation will be useful when we analyze the centroid
position below.

\subsubsection{Total magnification and centroid}

If the two images are not separately resolved, the main
observables are the total magnification and
magnification-weighted centroid position (e.g., \cite{GP}).
Applying the universal magnification relations
(\ref{eq:magreln}) to the total magnification yields
\beqa
  \mu_{\rm tot} &=& |\mu^+| +  |\mu^-| , \nonumber \\
  &=& (\mu_0^+ - \mu_0^-) \ + \ (\mu_1^+ - \mu_1^-)\,\vep
    \ + \ (\mu_2^+ - \mu_2^-)\,\vep^2 \ + \ \order{\vep}{3} , \nonumber \\ 
  &=& (2 \mu_0^+ - 1) \ + \ 2 \mu_2^+\,\vep^2 \ + \ \order{\vep}{3} .
    \label{eq:magtot}
\eeqa
There is no first-order correction to the total magnification
for all gravity theories within our PPN framework.  This
result depends on a precise cancellation between $\mu_1^+$
and $\mu_1^-$, so it is striking that it is universal for
PPN models.  An important implication is that the total
magnification would have to be measured much more precisely
than some other observables to find corrections to the
weak-deflection limit (see Sec.\ \ref{sec:applications} for
more discussion).

The magnification-weighted centroid position is defined by
\beq
  \Theta_{\rm cent} =  \frac{\theta^{+} |\mu^{+}| - \theta^{-} |\mu^{-}|}
    {|\mu^{+}| + |\mu^{-}|}
  = \frac{\theta^{+} \mu^{+} + \theta^{-} \mu^{-}}
    {\mu_{+} - \mu_{-}}\ .
\eeq
Writing $\theta^{\pm}$ and $\mu^{\pm}$ in terms of their series
expansions, and using the magnification relations (\ref{eq:magreln}),
yields
\beq
  \Theta_{\rm cent} = \Theta_0 \ + \ \Theta_1\,\vep \ + \ \Theta_2\,\vep^2
    \ + \ \order{\vep}{3} ,
\eeq
where
\beqa
  \Theta_0 &=& \frac{\theta_0^+ \mu_0^+ + \theta_0^- \mu_0^-}
    {\mu_0^+ - \mu_0^-}\ , \label{eq:Theta-0} \\
  \Theta_1 &=& \frac{ \mu_0^- \theta_1^- + 
    \mu^+_1 \theta_0^+ + \mu_1^- \theta_0^- + \mu_0^+ \theta_1^+ }
    { \mu_0^+ - \mu_0^- }\ , \label{eq:Theta-1} \\
  \Theta_2 &=& \frac{ 
    (\theta_0^+ + \theta_0^-)(\mu_0^+ \mu_2^- - \mu_0^- \mu_2^+)
    + (\mu_0^+ - \mu_0^-)[ \mu_1^+(\theta_1^+ + \theta_1^-)
      + \mu_0^+ \theta_2^+ + \mu_0^- \theta_2^- ] }{ (\mu_0^+ - \mu_0^-)^2 }\ .
    \label{eq:Theta-2}
\eeqa
Writing (\ref{eq:Theta-0}) and (\ref{eq:Theta-2}) in terms of
$\beta$ yields the zeroth-order centroid position and the
second-order correction to be
\beqa
  \Theta_0 &=& |\beta|\ \frac{3 A_1 + 4 \beta^2}{2 A_1 + 4 \beta^2}\ , \\
  \Theta_2 &=& - \frac{ |\beta| \{ 9 A_2^2 - 2 (A_1 + \beta^2)
    [ 6 A_3 + 8 A_1 D^2 \beta^4 - 6 A_1^2 D \beta^2 (2 - 3 D)
    + A_1^3 (2 - D^2) ] \} }{ 6 (A_1+\beta^2) (A_1+2\beta^2)^2 }\ .
\eeqa
These results are neither universal nor source-independent, but
are useful generalizations of previous results to the case
$A_1 \ne 4$.

The numerator of the first-order correction (\ref{eq:Theta-1}) is
identical to the left-hand side of (\ref{eq:posmag}), yielding
\beq \label{eq:centroid1-vanish}
  \Theta_1 = 0 .
\eeq
As with the total magnification, the first-order correction to
the centroid vanishes universally in the PPN framework,
which also means that centroid corrections beyond zeroth-order
will be more challenging to observe directly.

{\em Remark.}  For the special case of the Schwarzschild metric
in general relativity, Ebina et al.\ \cite{ebinaetal} and Lewis
\& Wang \cite{lewis-wang} found that the first-order corrections
to the total magnification and centroid vanish.  We have now
generalized that result to {\em all} PPN models.

\subsubsection{Differential time delay}

In many cases the only observable time delay is the differential
delay between the positive- and negative-parity images,
\beq
  \Delta\tauhat = \tauhat^- - \tauhat^+ ,
\eeq
which we write as a series of the form
\beq
  \Delta\tauhat = \Delta\tauhat_0 \ + \ \Delta\tauhat_1\,\vep
    \ + \ \order{\vep}{2} .
\eeq
Starting from (\ref{eq:tau})--(\ref{eq:tau1}) and using
(\ref{eq:theta0pm}) to write $\theta_0$ in terms of $\beta$,
we obtain
\beqa
  \Delta\tauhat_0 &=& \frac{1}{2}\, |\beta| \sqrt{A_1+\beta^2}
    \ + \ \frac{A_1}{4}\ \ln\left(
      \frac{\sqrt{A_1+\beta^2}+\beta}{\sqrt{A_1+\beta^2}-\beta} \right) \,,
    \label{eq:dt0} \\
  \Delta\tauhat_1 &=& \frac{A_2}{A_1}\ |\beta|\,. \label{eq:dt1}
\eeqa
These relations are not universal in the sense that each depends
on PPN parameters as well as the source position, but they are
still useful for calculations.

\section{Implications for Observations}
\label{sec:applications}

In this section we employ the new lensing relations to assess
prospects for using lensing observations to test theories of
gravity.  We consider three likely astrophysical scenarios:
lensing by the Galactic black hole; conventional Galactic
microlensing; and lensing in a binary pulsar system.

\subsection{Review of observables}

To facilitate this discussion, let us review the possible
lensing observables, focusing on the weak-deflection limits
plus the first-order corrections which should be measurable
now or in the near future.  To connect with realistic
observations, we revert from our convenient mathematical
variables ($\theta$, $\beta$, $\mu$, $\tauhat$) to true
observable quantities ($\vth$, $\cb$, $F$, $\tau$).

The traditional lensing observables are the positions, fluxes,
and time delays of the images.  The fluxes are related to the
magnifications via the source flux: $F_i = |\mu_i|\,F_{\rm src}$.
(As an observable quantity, flux is positive-definite.)  In
principle, one could simply take the measured values, adopt
the formulas derived in this paper as a model, and fit for
the unknown parameters.  However, with the relations found
in Section \ref{sec:relations} as a guide, we believe it is
instructive to make certain combinations of observables as
follows:
\beqa
  \vth^+ \ + \ \vth^- &=& \sqrt{A_1\vthE^2+\cb^2}
    \ + \ \frac{A_2\,\vthE}{A_1}\ \vep \ + \ \order{\vep}{2} ,
    \label{eq:pos+} \\
  \vth^+ \ - \ \vth^- &=& |\cb|
    \ - \ \frac{A_2\,\vthE\,|\cb|}{A_1 \sqrt{A_1\vthE^2+\cb^2}}\ \vep
    \ + \ \order{\vep}{2} ,
    \label{eq:pos-} \\
  F_{\rm tot} \equiv
  F^+ \ + \ F^- &=& F_{\rm src}\ \frac{A_1\vthE^2 + 2\cb^2}
    {2 |\cb| \sqrt{A_1\vthE^2+\cb^2}} \ + \ \order{\vep}{2} ,
    \label{eq:flux+} \\
  \Delta F \equiv
  F^+ \ - \ F^- &=& F_{\rm src}
    \ - \ F_{\rm src}\ \frac{A_2\,\vthE^3}{2 (A_1\vthE^2+\cb^2)^{3/2}}\ \vep
    \ + \ \order{\vep}{2} ,
    \label{eq:flux-} \\
  \vartheta_{\rm cent} \equiv
  \frac{\vth^+\, F^+ - \vth^- \, F^-}{F_{\rm tot}} &=& 
  |\cb|\ \frac{3 A_1\vthE^2 + 4 \cb^2}
    {2 A_1\vthE^2 + 4 \cb^2} \ + \ \order{\vep}{2} ,
    \label{eq:cent} \\
  \Delta\tau &=& \frac{d_L\,d_S}{c\,d_{LS}} 
    \Biggl[ \frac{1}{2}\, |\cb| \sqrt{A_1\vthE^2+\cb^2}
      \ + \ \frac{A_1\,\vthE^2}{4}\ \ln\left(
        \frac{\sqrt{A_1\vthE^2+\cb^2}+|\cb|}{\sqrt{A_1\vthE^2+\cb^2}-|\cb|}
        \right) \nonumber\\
  &&\qquad\qquad
      \ + \ \frac{A_2\,\vthE}{A_1}\ |\cb|\,\vep
      \ + \ \order{\vep}{2} \Biggr] . \label{eq:tdel}
\eeqa
These equations summarize our results for PPN lensing, and
play the key role in understanding how lensing can be used
to test PPN theories of gravity.

\subsection{The Galactic black hole}

The center of our Galaxy is believed to host a supermassive
black hole with a mass of
$\Mbh = (3.6\pm0.2) \E{6}\,M_\odot$ \cite{ghez}; the distance
to the lens is $d_L = 7.9 \pm 0.4$ kpc \cite{eisen}.  Adopting
the nominal values and neglecting the small uncertainties,
we find the black hole's gravitational radius to be
$\gravr = 5.3\E{9}\mbox{ m} = 1.7\E{-7}\mbox{ pc}$, which
corresponds to an angle of $\vthbh = 4.5\E{-6}$ arc s.  The
corresponding lensing time scale is $\tau_E = 71$ s.

We consider a source that is orbiting the black hole at a
distance $d_{LS} \ll d_L$ (so $d_S \approx d_L$).  In the
following quantitative estimates we let
$d^*_{LS} = d_{LS}/(1\mbox{ pc})$ to simplify the notation.
If the orbit is close to edge-on, part of it will lie close
enough to the black hole (in projection) that the source can
be significantly lensed.  The angular Einstein radius is then
$\vthE = 0.022\,(d^*_{LS})^{1/2}$ arc s, and our dimensionless
perturbation parameter is
$\vep = 2.1\E{-4} \times (d^*_{LS})^{-1/2}$.

These numbers indicate that the two lensed images could be
resolved with existing technology.  From (\ref{eq:pos+}), the
angular separation of the two images is at least
$\sqrt{A_1}\ \vthE = 2\,\vthE = 0.044\,(d^*_{LS})^{1/2}$ arc s
(using $A_1 = 3.99966\pm0.00090$ \cite{sh}).  At optical
wavelengths, the Hubble Space Telescope has a resolution of
about 0.05 arc s, while the CHARA interferometer \cite{CHARA}
can obtain a resolution of better than $10^{-3}$ arc s.  At
radio wavelengths, interferometry can also achieve a resolution
of $\sim\!10^{-3}$ arc s.  The position uncertainties are
much smaller than the resolution element; for example, radio
interferometry observations of known lenses have yielded
position uncertainties at the level of $10^{-6}$ arc s
(e.g., \cite{trotter}).  With the conservative assumption
that the images must be separated by $10^{-2}$ arc s to be
well resolved, we can still consider sources as close to the
black hole as $d_{LS} = 0.05$ parsec, and we can expect to
measure the image positions with micro-arcsecond precision.

A single observation could therefore be expected to yield the
position and flux of each of the two images.  Using equations
(\ref{eq:pos+})--(\ref{eq:flux-}), the four numbers 
($\vartheta^+$, $\vartheta^-$, $F^+$, $F^-$) would allow us
to solve for four unknowns, which we could take to be $d_{LS}$,
$\cb$, $F_{\rm src}$, and $A_2$.  (Here we imagine taking the
values of $\Mbh$, $d_L$, and $A_1$ as given above.)
Thus, in principle a single observation of an appropriate
source could test gravity theories by measuring the invariant
PPN bending-angle parameter $A_2$.  The position shifts that
depend on $A_2$ are of order
$\vthE\,\vep = \vthbh = 4.5\E{-6}$ arc s, so existing
technology has sufficient precision to measure them.  Note
that $A_2$ is connected to the occurrence of naked singularities
for certain gravity theories (see Sec.\ III.D of Paper I), so
lensing could provide an observational test of the Cosmic
Censorship conjecture \cite{CosmicCensorship}.  This analysis
indicates that the main challenge for using lensing by the
Galactic black hole to test gravity theories is just to find
a source that is lensed.

We could do even better with repeated observations, watching
as the source moves and causes the images to change.  We may
estimate the time scale for such variations as the time it
takes the source to move one linear Einstein radius,
$d_S\,\vth_E$.  For a circular orbit, this is
$T_E = 6.5\,d^*_{LS}$ yr (independent of the black hole mass,
since the Einstein radius and the Keplerian orbital velocity
both scale as $\Mbh^{1/2}$).  Keplerian orbital motion can be
described with just five parameters: semimajor axis, period,
eccentricity, inclination, and longitude of periastron.
Repeated observations would thus allow us to determine a
good model for $\cb$ as a function of time.  The universal
relations then tell us that the quantity $\vth^+ - \vth^-$
would be the most interesting combination of observables.
{\em From (\ref{eq:pos-}), if $\vth^+ - \vth^-$ has any dependence
on $\cb$ that is not strictly linear, that would represent a
clear detection of higher-order effects from the gravity
theory.}  The practical value of the universal relations,
then, is to identify combinations of observables that would
give the most direct evidence that the measurements are
probing beyond the weak-deflection limit.

The other reason to make repeated observations is of course
to obtain more observables than unknowns, so the problem
becomes overconstrained.  In this case the data will not
only determine the parameter values, but determine whether
the PPN framework itself is an acceptable model.

\subsection{Galactic microlensing}

In conventional microlensing, a foreground star or compact
object lenses a star in the Galactic bulge.  In what follows
we quote $\Mbh$ in units of the mass of the Sun, and $d_S$
in units of 8 kpc: $\Mbh^* = \Mbh/M_\odot$ and
$d^*_S = d_S/(8\, {\rm kpc})$.  We consider a typical
situation with the lens lying about halfway between the
observer and source: $d_L \sim d_{LS} \sim d_S/2$.
The angular gravitational radius is then
$\vthbh \sim 2.5\E{-12} \times (\Mbh^*/d^*_S)$ arc s,
the angular Einstein radius is
$\vthE \sim 10^{-3}\,(\Mbh^*/d^*_S)^{1/2}$ arc s,
and the perturbation parameter is
$\vep \sim 2.4\E{-9} \times (\Mbh^*/d^*_S)^{1/2}$.

Present microlensing programs only measure the total flux
as a function of time (or equivalently source position).
Equation (\ref{eq:flux+}) shows that there is no
first-order correction to the total flux, so it is not
feasible to test theories of gravity with microlensing at
present.  Future programs may be able to resolve the images
(which will be separated by a few milli-arcseconds).  But
in order to test gravity theories they would need to measure
image positions with a precision at the level of
$10^{-12}$ arc s, or fluxes with a fractional uncertainty
of order $\vep \sim 10^{-9}$.  In other words, it is not
reasonable to expect to test theories of gravity with
conventional microlensing in the foreseeable future.

\subsection{Pulsars in binary systems}

Hopes for using stellar-mass lenses to test theories of
gravity are not lost.  The amplitudes of the correction
terms are governed by
\beq
  \vep = \left( \frac{G \Mbh}{4 c^2}\ \frac{d_S}{d_L\,d_{LS}} \right)^{1/2} ,
\eeq
so we may be able to use stellar-mass lenses if we can find
systems where $d_{LS}$ is sufficiently small.  The ideal
system would be a pulsar in a binary system with a compact
object (another pulsar, or a black hole), in an orbit seen
nearly edge-on.  An example of such a system was recently
discovered: the binary pulsar J0737$-$3039 \cite{j0737}.
Rafikov \& Lai \cite{rafikov} have made detailed calculations
of various effects on the pulsar timing measurements,
including not only multiple imaging but also relativistic
aberration and latitudinal delays associated with the spin
of the source.  We use this system more generally to be
representative of binary systems consisting of a pulsar and
a compact object, and to illustrate the amplitude of lensing
effects associated with different theories of gravity.

In J0737$-$3039, we take the fast millisecond pulsar to be
the light source, and the slow pulsar with $\Mbh = 1.25\,M_\odot$
to be the lens.  The binary orbit has a semimajor axis
$a = 8.78\E{5}$ km.  The orbital eccentricity is fairly small
($e = 0.088$), so for illustration purposes take $d_{LS} = a$.
The lens gravitational radius is then $\gravr = 1.8$ km, so
the lensing time scale is $\tau_E = 2.5\E{-5}$ s.  
Using $d_L \approx d_S$, the physical
Einstein radius is
$d_L\,\vthE \approx (4 G \Mbh d_{LS}/c^2)^{1/2} = 2.5\E{3}$ km.
(The angular Einstein radius cannot be determined because the
distance to the lens is not known.)  The perturbation parameter
is $\vep = 7.2\E{-4}$.

The two images could not be resolved spatially.  They could
be resolved temporally, though: when the source is behind the
lens, each radio pulse would actually consist of two pulses
(one from each image) separated by the lens time delay
(see \cite{rafikov}).  The amplitudes of the two pulses
could be measured; and the intrinsic pulse amplitude could
be measured when the source is not behind the lens.  Thus,
in this system the observables would be $F^+$, $F^-$, and
$\Delta\tau$ as a function of source position $\cb$, as well
as $F_{\rm src}$.  Should one wish to go further, analysis
of the source's orbital motion would yield a prediction for
the pulse arrival time in the absence of lensing and make it
possible to measure the time delays $\tau^\pm$ for the two
images separately.

In this scenario, the universal relations indicate that for
testing gravity theories the most valuable measurement would
be the flux difference $\Delta F = F^+ - F^-$ as a function
of source position.  From (\ref{eq:flux-}), this quantity is 
constant in the weak-deflection limit.  Thus, {\em any
variation in $\Delta F$ with source position would reveal
that the measurements are probing beyond the weak-deflection
limit}.  Again we see the universal relations helping us
identify combinations of observables that are best suited
for testing gravity theories.

\section{Conclusions}
\label{sec:conclusions}

Using gravitational lensing by compact objects, we have presented
new propects for testing theories of gravity within the PPN
framework.  In this paper we generalized the PPN lensing
formalism from Paper I to include fully general third-order
PPN models.  We determined the weak-deflection limits plus
first- and second-order corrections in $\vep = \vthbh/\vthE$
for observable properties of lensed images (positions,
magnifications, and time delays).

During the PPN analysis, we discovered some surprising new
fundamental relations between lensing observables.  Some of
the relations are universal for the entire family of PPN
gravity models.  A deep conceptual implication is that any
observed violation of the universal lensing relations (given
that assumptions \assump{1}--\assump{3} apply) would indicate
that a fundamentally different theory of gravity is at work
--- one outside the PPN framework.  The new relations have
enabled us to identify combinations of lensing observables
that are key to probing the spacetime metric by constraining
the invariant PPN parameters of the light bending angle.
The parameter $A_2$ is related to the existence of naked
singularities in certain gravity models (see Paper I), so
constraining $A_2$ also provides a possible observational
test of the Cosmic Censorship conjecture.  The new lensing
relations will, in other words, play important roles in
planning observing missions to test theories of gravity.  

In a practical application, we identified lensing
observables that are accessible to current or near-future
instrumentation, considering three likely lensing scenarios:
the Galactic black hole, Galactic microlensing, and the
binary pulsar J0737$-$3039.  A noted application of the
new lensing relations is the ability to find combinations
of observables that will yield a direct method for knowing
when observations are probing beyond the standard
weak-deflection regime.

\begin{acknowledgments}

This work was supported by NSF grants DMS-0302812, AST-0434277,
and AST-0433809.  

\end{acknowledgments}

\end{document}